\documentclass[12pt]{article}
\usepackage{fancyhdr}

\usepackage{mathrsfs}
\usepackage[T1]{fontenc}
\usepackage{mathpazo}
\usepackage{setspace}
\usepackage{amsfonts}
\usepackage{amssymb}
\usepackage{amsmath}
\usepackage{epsfig}
\usepackage{latexsym}
\usepackage{color}
\usepackage{graphicx}
\usepackage{nicefrac}
\usepackage[latin1]{inputenc}
\usepackage{slashed}
\usepackage{multirow}
\usepackage{comment}
\usepackage{soul}
\usepackage{hyperref}
\usepackage{cite}
\usepackage{datetime}

\usepackage[titletoc]{appendix}

\def\hybrid{\topmargin -20pt    \oddsidemargin 0pt
        \headheight 0pt \headsep 0pt
        \textwidth 6.25in       
        \textheight 9.25in       
        \marginparwidth .875in
        \parskip 5pt plus 1pt   \jot = 1.5ex}

\hybrid

\def\baselinestretch{1.2}

\catcode`\@=11

\def\marginnote#1{}
\newcount\hour
\newcount\minute
\newtoks\amorpm
\hour=\time\divide\hour by60
\minute=\time{\multiply\hour by60 \global\advance\minute by-\hour}
\edef\standardtime{{\ifnum\hour<12 \global\amorpm={am}%
        \else\global\amorpm={pm}\advance\hour by-12 \fi
        \ifnum\hour=0 \hour=12 \fi
        \number\hour:\ifnum\minute<10 0\fi\number\minute\the\amorpm}}
\edef\militarytime{\number\hour:\ifnum\minute<10 0\fi\number\minute}

\def\draftlabel#1{{\@bsphack\if@filesw {\let\thepage\relax
   \xdef\@gtempa{\write\@auxout{\string
      \newlabel{#1}{{\@currentlabel}{\thepage}}}}}\@gtempa
   \if@nobreak \ifvmode\nobreak\fi\fi\fi\@esphack}
        \gdef\@eqnlabel{#1}}
\def\@eqnlabel{}
\def\@vacuum{}
\def\draftmarginnote#1{\marginpar{\raggedright\scriptsize\tt#1}}

\def\draft{\oddsidemargin -.5truein
        \def\@oddfoot{\sl preliminary draft \hfil
        \rm\thepage\hfil\sl\today\quad\militarytime}
        \let\@evenfoot\@oddfoot \overfullrule 3pt
        \let\label=\draftlabel
        \let\marginnote=\draftmarginnote
   \def\@eqnnum{(\theequation)\rlap{\kern\marginparsep\tt\@eqnlabel}%
\global\let\@eqnlabel\@vacuum}  }

\def\preprint{\twocolumn\sloppy\flushbottom\parindent 2em
        \leftmargini 2em\leftmarginv .5em\leftmarginvi .5em
        \oddsidemargin -.5in    \evensidemargin -.5in
        \columnsep .4in \footheight 0pt
        \textwidth 10.in        \topmargin  -.4in
        \headheight 12pt \topskip .4in
        \textheight 6.9in \footskip 0pt
        \def\@oddhead{\thepage\hfil\addtocounter{page}{1}\thepage}
        \let\@evenhead\@oddhead \def\@oddfoot{} \def\@evenfoot{} }

\def\numberbysection{\@addtoreset{equation}{section}
        \def\theequation{\thesection.\arabic{equation}}}

\def\underline#1{\relax\ifmmode\@@underline#1\else
        $\@@underline{\hbox{#1}}$\relax\fi}

\def\titlepage{\@restonecolfalse\if@twocolumn\@restonecoltrue\onecolumn
     \else \newpage \fi \thispagestyle{empty}\c@page\z@
        \def\thefootnote{\fnsymbol{footnote}} }

\def\endtitlepage{\if@restonecol\twocolumn \else \newpage \fi
        \def\thefootnote{\arabic{footnote}}
        \setcounter{footnote}{0}}  

\catcode`@=12
\relax

\def\figcap{\section*{Figure Captions\markboth
        {FIGURECAPTIONS}{FIGURECAPTIONS}}\list
        {Figure \arabic{enumi}:\hfill}{\settowidth\labelwidth{Figure
999:}
        \leftmargin\labelwidth
        \advance\leftmargin\labelsep\usecounter{enumi}}}
 \relax
\def\tablecap{\section*{Table Captions\markboth
        {TABLECAPTIONS}{TABLECAPTIONS}}\list
        {Table \arabic{enumi}:\hfill}{\settowidth\labelwidth{Table
999:}
        \leftmargin\labelwidth
        \advance\leftmargin\labelsep\usecounter{enumi}}}
 \relax
\def\reflist{\section*{References\markboth
        {REFLIST}{REFLIST}}\list
        {[\arabic{enumi}]\hfill}{\settowidth\labelwidth{[999]}
        \leftmargin\labelwidth
        \advance\leftmargin\labelsep\usecounter{enumi}}}
 \relax
\makeatletter
\newcounter{pubctr}
\def\publist{\@ifnextchar[{\@publist}{\@@publist}}
\def\@publist[#1]{\list
        {[\arabic{pubctr}]\hfill}{\settowidth\labelwidth{[999]}
        \leftmargin\labelwidth
        \advance\leftmargin\labelsep
        \@nmbrlisttrue\def\@listctr{pubctr}
        \setcounter{pubctr}{#1}\addtocounter{pubctr}{-1}}}
\def\@@publist{\list
        {[\arabic{pubctr}]\hfill}{\settowidth\labelwidth{[999]}
        \leftmargin\labelwidth
        \advance\leftmargin\labelsep
        \@nmbrlisttrue\def\@listctr{pubctr}}}
 \relax
\makeatother
\newskip\humongous \humongous=0pt plus 1000pt minus 1000pt

\newif\ifdtup

\relax

\def\be{\begin{equation}}
\def\ee{\end{equation}}
\def\ba{\begin{eqnarray}}
\def\ea{\end{eqnarray}}

\def\del{\partial}

\def\k{\kappa}

\def\a{\alpha}

\def\b{\beta}

\def\g{\gamma}

\def\d{\delta}
\def\D{\Delta}

\def\m{\mu}

\def\l{\lambda}
\def\L{\Lambda}
\def\s{\sigma}
\def\S{\Sigma}

\def\no{\noindent}

\def\qq{\qquad}

\def\IR{\relax{\rm I\kern-.18em R}}

\def \ov {\over}

\def\IR{\relax{\rm I\kern-.18em R}}
\def\IL{\relax{\rm I\kern-.18em L}}

\def\inv{^{\raise.15ex\hbox{${\scriptscriptstyle -}$}\kern-.05em 1}}

\def\Tr{{\rm Tr}}

\begin{document}

\renewcommand{\theequation}{\thesection.\arabic{equation}}
\csname @addtoreset\endcsname{equation}{section}

\newcommand{\beq}{\begin{equation}}
\newcommand{\eeq}[1]{\label{#1}\end{equation}}
\newcommand{\ber}{\begin{equation}}
\newcommand{\eer}[1]{\label{#1}\end{equation}}
\newcommand{\eqn}[1]{(\ref{#1})}
\begin{titlepage}
\begin{center}

\hfill CERN-TH-2017-199

${}$
\vskip .2 in

{\large\bf Integrable deformations of the $G_{k_1} \times G_{k_2}/G_{k_1+k_2}$ coset CFTs}

\vskip 0.4in

{\bf Konstantinos Sfetsos}$^{1,2}$\ \ and\ \ {\bf Konstantinos Siampos}$^{2,3}$
\vskip 0.1in

 {\em
${}^1$Department of Nuclear and Particle Physics,\\
Faculty of Physics, National and Kapodistrian University of Athens,\\
15784 Athens, Greece\\
}

\vskip 0.1in
 {\em
${}^2$Theoretical Physics Department,\\
CERN, CH-1211 Geneva 23, Switzerland
}
\vskip 0.1in

{\em${}^3$Albert Einstein Center for Fundamental Physics,\\
Institute for Theoretical Physics,
University of Bern,\\
Sidlerstrasse 5, CH3012 Bern, Switzerland
}

\vskip 0.1in

{\footnotesize \texttt ksfetsos@phys.uoa.gr, konstantinos.siampos@cern.ch}


\vskip .5in
\end{center}

\centerline{\bf Abstract}

\no
We study the effective action for the integrable $\lambda$-deformation of the $G_{k_1} \times G_{k_2}/G_{k_1+k_2}$ coset CFTs.
For unequal levels theses models do not fall into the general discussion of $\lambda$-deformations of CFTs corresponding to
symmetric spaces and have many attractive features. We show that the perturbation is driven by parafermion
bilinears and we revisit the derivation of their algebra. We uncover a non-trivial symmetry of these models parametric space, 
which has not encountered before in the literature. Using field theoretical methods and the effective action we compute the exact in the deformation parameter $\beta$-function
and explicitly demonstrate the existence of a fixed point in the IR corresponding to the
 $G_{k_1-k_2} \times G_{k_2}/G_{k_1}$ coset CFTs.  The same result is verified using gravitational methods for $G=SU(2)$.
We examine various limiting cases previously considered in the literature and found agreement.

\vskip .4in
\noindent
\end{titlepage}
\vfill
\eject

\newpage

\tableofcontents

\noindent

\def\baselinestretch{1.2}
\baselineskip 20 pt
\noindent


\setcounter{equation}{0}

\section{Introduction}

Perturbing a conformal field theory (CFT) while maintaining integrability,  especially to all perturbative orders, usually proves quiet challenging.
In this paper we consider the $G_{k_1}\times G_{k_2}/G_{k_1+k_2}$  coset CFTs, where $G$
is a semi-simple group and the levels $k_1,k_2$ are generically  different.  
Then we perturb these CFTs
using bilinears of operators {with} conformal dimension
\begin{equation}
\label{dk1k2}
\D_{k_1,k_2}= 1-{c_G\ov 2( k_1+k_2)+ c_G}\ ,
\end{equation}
where $c_G$ is the quadratic Casimir in the adjoint representation of $G$.
These models were extensively studied in the past \cite{Bernard:1990cw,Crnkovic:1989gy,Ahn1990,Zamolodchikov:1991vg,Ravani},
mostly for $G=SU(2)$.
In particular, it has been argued based on thermodynamic Bethe ansatz (TBA) considerations and on perturbative computations
when one of the levels is much larger than the other one, that there is an integrable flow from the
$G_{k_1}\times G_{k_2}/G_{k_1+k_2}$ coset CFTs, in the UV, to the
$G_{k_1-k_2}\times G_{k_2}/G_{k_1}$ coset CFTs in the IR. Moreover, it was also argued
that these models have interesting limits when one or both levels are taken to infinity.

\no
In the present work we make considerable progress along the above research line by constructing an effective action for these
models valid to all orders in the perturbation parameter. The latter will be alternatively called deformation parameter, emphasizing the
non-trivial dependence the action will have on it. This effective action will be nothing but that the
$\l$-deformed coset $G_{k_1} \times G_{k_2}/G_{k_1+k_2}$ coset which will be used to
explicitly prove the previous properties and more.
The construction will follow the rules of the usual (integrable) $\l$-deformations for current algebras $G_k$ \cite{Sfetsos:2013wia} and
for $G_k/H_k$ (symmetric) coset CFTs \cite{Sfetsos:2013wia,Hollowood:2014rla},
appropriately generalized to take into account the presence of two different levels \cite{Sfetsos:2014cea}.
As is the case for all $\l$-deformed type actions \cite{Sfetsos:2013wia,Georgiou:2016urf,Georgiou:2017jfi,Georgiou:2017oly,Georgiou:2016zyo}
the resulting action will be valid for large levels but exact in the deformation parameter. 

The structure of this work is as follows. In section \ref{action.section}, we explicitly construct the effective action for the $\l$-deformed
$G_{k_1}\times G_{k_2}/G_{k_1+k_2}$ coset CFTs. Then we show that they possess the non-trivial symmetry \eqref{symmetry}
in their parametric space $(\l,k_1,k_2)$ and prove their classical integrability by rewriting their
equations of motion in the Lax form \eqref{integrabb}. In section \ref{limits.section}, we study various limits
when the levels $k_1$ and/or $k_2$ are taken to infinity and we make connection with related statements in
the literature \cite{Ahn1990,Zamolodchikov:1991vg,Ravani}. In addition, we discover a new limit
which is the non-Abelian transformation of the $\l$-deformed WZW model for a group $G$.
In section \ref{RGflows.section}, we compute the exact
$\beta$-function in the deformation parameter  \eqref{betafunction} and the IR fixed point, 
which has been argued to exist previously. In addition, we study its various properties and limits.
For unequal levels, this $\beta$-function is not what one obtains for the $\l$-deformed coset models for symmetric spaces.
This is explained by the non-Abelian nature of the parafermionic algebra \eqref{paraalgebra} for the $G_{k_1}\times G_{k_2}/G_{k_1+k_2}$ coset CFTs whose derivation is revisited in the appendix \ref{Para.append} but
originally derived in \cite{Bardakci:1990ad}.

\section{The effective action and its integrability}
\label{action.section}

The effective action will be constructed by following the rules in \cite{Sfetsos:2013wia} as extended for the coset models in question in \cite{Sfetsos:2014cea}. 
Hence, we consider a sum of WZW actions for the group elements $g_1, g_2\in G$,
at different levels $k_1$ and $k_2$ and add to them the principal chiral model (PCM) action
for the coset $G\times G/ G$ with some overall coupling constant
$\k^2$.
Subsequently one gauges the subgroup $G$ acting vectorially on $g_1$ and $g_2$, and from the left on the group
elements on the coset PCM. In order to make the action gauge invariant,
we introduce gauge fields $A_{1\pm}$ and $A_{2\pm}$ in the Lie algebras
of the group $G\times G$. Since the gauge group's action is free on the PCM group elements, we can fix them to unity.
After this gauge fixing the contribution of the PCM is simply
$$ -{2\k^2\ov \pi} \int  \text{d}^2\s\ {\rm Tr} \left(\mathcal{B}_+\mathcal{B}_-\right)\ , \quad
\mathcal{B}_\pm=\frac12\left(A_{1\pm}-A_{2\pm}\right)\ ,$$
where the minus relative sign between the gauge fields
is simply due to the fact that the coset generators correspond to the difference of the generators of the two
groups in $G \times G$ and the subgroup to their sum.
Then the total action is
\begin{equation}
\begin{split}
\label{acct1}
&
S_{k,\k^2}(g,A_{1,2\pm}) = \sum_{i=1}^2\left\{S_{k_i}(g_i)
 +{k_i\ov \pi} \int \text{d}^2\s\ \Tr \big(A_{i-} \del_+ g_i g_i^{-1} - A_{i+} g_i^{-1} \del_- g_i\right.
 \\
 &\left.\phantom{x} + A_{i-} g_i A_{i+} g_i^{-1}  -A_{i+} A_{i-} \big)\right\}
-k {\l^{-1}-1\ov \pi} \int \text{d}^2\s\ {\rm Tr}\left(\mathcal{B}_+\mathcal{B}_-\right) \ ,
\end{split}
\end{equation}
where the WZW action for a group $G$ is
\begin{equation*}
S_k(g)= {k\ov 2\pi} \int \text{d}^2\s\, \Tr(\del_+ g^{-1} \del_- g)
+ {k\ov 12\pi} \int  \Tr(g^{-1} \text{d}g)^3\
\end{equation*}
and we found it convenient to introduce the parameters
\begin{equation}
\label{ldef}
\l = {k\ov k+ 2 \k^2}\ , \quad k=k_1+k_2\ ,\quad s_i ={k_i\ov k}\ ,\quad i=1,2\ .
\end{equation}
We note that the gauge freedom should be completely fixed by choosing additional $\dim G$-parameters in the groups elements $g_1$ and
$g_2$, therefore leaving $\dim G$ group parameters in total. These will be the background coordinates in the $\s$-model
action to be derived. This gauge fixing has to be done on a case by case basis, depending on the specific parametrization 
of the group elements. 

Next we find the equations of motion for the above action.
Varying \eqref{acct1} with respect to the $A_{i\pm}$'s, we find the following constraints
\begin{equation}
\begin{split}
s_1 D_+ g_1\, g_1^{-1} =  {1\ov 2} (\l^{-1}-1) \mathcal{B}_+\ ,\quad
  s_2 D_+ g_2\, g_2^{-1} = - {1\ov 2} (\l^{-1}-1) \mathcal{B}_+\
\label{dggd}
\end{split}
\end{equation}
and
\begin{equation}
\begin{split}
s_1 g^{-1}_1 D_- g_1 =-{1\ov 2} (\l^{-1}-1) \mathcal{B}_- \ ,\quad
s_2 g^{-1}_2 D_- g_2 ={1\ov 2} (\l^{-1}-1) \mathcal{B}_- \ .
\label{dggd2}
\end{split}
\end{equation}
Note that from these it follows that
\begin{equation}
s_1  D_+ g_1\, g_1^{-1} + s_2 D_+ g_2\, g_2^{-1}  = 0 \ ,\qq s_1 g^{-1}_1 D_- g_1   + s_2 g^{-1}_2 D_- g_2 =0\ ,
\label{sumsum}
\end{equation}
implying, that the gauge invariant subgroup current for the left and the right chiralities vanishes on-shell, as it should.
Returning to  the equations of motion, varying the action with respect to group elements $g_1$ and $g_2$ results into
\begin{equation}
\label{eqg1g2}
D_ -(D_+ g_i g_i^{-1})= F_{i +-}\ ,\quad i=1,2\ ,
\end{equation}
where
\begin{equation*}
F_{i+-}=\del_+ A_{i-} - \del_- A_{i+} - [A_{i+},A_{i-}]\ , \quad i=1,2\ .
\end{equation*}
Equivalently, these can be written as
\begin{equation}
D_+(g_i^{-1}D_- g_i)= F_{i+-}\ , \quad i=1,2\ .
\label{eqg1g22}
\end{equation}
The covariant derivatives are defined according to the group element contained in
the object on which they act. For example, $D_\pm g_1= \del_\pm g_1 -[A_{1\pm}, g_1]$.
Next we define
\begin{equation}
\label{hg3}
J^a_+ = - i\, \Tr(t_a \del_+ g g^{-1})\ ,\quad J^a_- = - i\,\Tr(t_a g^{-1}\del_- g )\ ,\quad
\quad D_{ab}= \Tr(t_a g t_b g^{-1})\ ,
\end{equation}
where $t_a$'s are Hermitian representation matrices obeying $[t_a,t_b]=if_{abc}t_c$, for
real structure constants $f_{abc}$. We choose the normalization such that $\Tr(t_at_b)=\d_{ab}$.
In what follows these quantities will have an extra index $1$ or $2$ depending on whether the group element $g_1$ or $g_2$
has been used.

In order to obtain the desired $\sigma$-model action we should integrate out the gauge fields which appear non-dynamically in
the system of equations \eqref{dggd} and \eqref{dggd2}. After some algebraic manipulations we find that
\begin{equation}
\label{eqpl}
\begin{split}
& A_{1+}=i \L_{21}^{-1}\big((1-\l)(s_1J_{1+}+ s_2J_{2+})- 4 s_1s_2 \l(D_2-\mathbb{I})J_{1+}\big)\ ,
\\
&
A_{2+}=i \L_{12}^{-1}\big((1-\l)(s_1J_{1+}+ s_2J_{2+})- 4 s_1s_2 \l(D_1-\mathbb{I})J_{2+}\big)\ 
\end{split}
\end{equation}
and
\begin{equation}
\label{eqmi}
\begin{split}
& A_{1-}= - i \L_{12}^{-T}\big((1-\l)(s_1J_{1-}+ s_2J_{2-})- 4 s_1s_2 \l(D_2^T-\mathbb{I})J_{1-}\big)\ ,
\\
&
A_{2-}= - i \L_{21}^{-T}\big((1-\l)(s_1J_{1-}+ s_2J_{2-})- 4 s_1s_2 \l(D_1^T-\mathbb{I})J_{2-}\big)\ ,
\end{split}
\end{equation}
where
\begin{equation}
\label{l12}
\L_{12}=4 \l s_1 s_2 (D_1-\mathbb{I})(D_2-\mathbb{I}) + (\l-1)\big(s_1D_1+s_2D_2-\mathbb{I})\big)\ ,
\end{equation}
with $\L_{21}$ following by interchanging the indices  $1$ and $2$.
Substituting these expressions into \eqref{acct1} results into the  $\sigma$-model action
\begin{equation}
\label{resact}
\begin{split}
& S_{k,\l}(g_1,g_2)= S_{k_1}(g_1) + S_{k_2}(g_2)
\\
& \phantom{xxxx}
+ {k\ov \pi} \int \text{d}^2\s \Big\{  s_1 J_{1+} \L_{12}^{-T} \left((1-\l)(s_1J_{1-}+ s_2J_{2-})- 4 s_1s_2 \l(D_2^T-\mathbb{I})J_{1-}\right)
\\
&
\phantom{xxxxxxxxx} + s_2 J_{2+} \L_{21}^{-T} \left((1-\l)(s_1J_{1-}+ s_2J_{2-})-
4 s_1s_2 \l(D_1^T-\mathbb{I})J_{2-}\right) \Big\} \ .
\end{split}
\end{equation}
For $\l\to 0$, the action has obviously a smooth limit given by
\be
\label{cftww}
\begin{split}
& S_{\rm CFT} =  S_{k_1}(g_1) + S_{k_2}(g_2)
\\
&  \phantom{xxxx} + {1\ov \pi} \int \text{d}^2\s (k_1 J_{1+} + k_2 J_{2+})(k\mathbb{I}-k_1D_1^T - k_2 D_2^T)^{-1}
 (  k_1 J_{1-} + k_2 J_{2-}) \ ,
\end{split}
\ee
which is the $\sigma$-model action
corresponding to the $G_{k_1}\times G_{k_2}/ G_{k_1+k_2}$  coset CFTs.

 It is very important to recognize the operator that drives the theory away from the
CFT point. To do so we should compute the ${\cal O}(\l)$ correction to the $S_{\rm CFT}$. Apparently, this computation cannot be performed very easily using \eqref{resact}.  Instead one may use \eqref{acct1}. We easily see that for $\l=0$, we have that  $\L_{12}=\L_{21}$ and $A_{1\pm}=A_{2\pm}$. Expanding the gauge fields as
$A_{1,2\pm }=A_{\pm}^{(0)} + \l A_{1,2\pm}^{(1)}+\dots $ and using this in \eqref{sumsum}, leads to
\begin{equation}
S_{k,\l}(g_1,g_2)= S_{\rm CFT}(g_1,g_2)+ 4\l {k\ov \pi} s_1 s_2 \int \text{d}^2\s\ {\rm Tr} \big(D_+^{(0)} g_1 g_1^{-1} g_2^{-1}
D_-^{(0)} g_2\big) +  \cdots \ ,
\end{equation}
where the superscript in the covariant derivative implies that the leading order expression for the gauge fields $A_\pm^{(0)}$
has been used, as the $A_{\pm}^{(1)}$ drops out completely to linear order in $\l$.
We may interpret this expression if we first rewrite it in a more suggestive form.
We define
\begin{equation}
\label{defparaf}
\begin{split}
&\Psi_+ = {1\ov 2}(s_1 D^{(0)}_+ g_1 g_1^{-1} - s_2 D^{(0)}_+ g_2 g_2^{-1}) \ ,
\\
&\Psi_- = -{1\ov 2}(s_1 g_1^{-1} D^{(0)}_- g_1 - s_2 g_2^{-1} D^{(0}_- g_2 ) \ .
\end{split}
\end{equation}
Then we easily see that, upon using \eqref{sumsum} for $\l=0$, the above perturbative expansion becomes
\begin{equation}
S_{k,\l}(g_1,g_2)= S_{\rm CFT}(g_1,g_2)+ 4\l {k\ov \pi}  \int \text{d}^2\s\ {\rm Tr} \big(\Psi_+ \Psi_-\big) +  \cdots \ .
\label{paraf}
\end{equation}
It has been shown quite generally \cite{Bardakci:1990ad,Bardacki:1990wj} that, by including Wilson lines, $\Psi_+$
and $\Psi_-$ as defined above, are chiral and
anti-chiral, respectively and become the classical non-Abelian parafermions \cite{Bardacki:1990wj}.
Due to the Wilson lines these are non-local objects and they have non-trivial monodromy properties. 
The Wilson lines attached to them drop out due to the
 fact that they appear within a trace so that the perturbation is eventually local as it should be.
 This  parafermion bilinear drives the model away from the conformal point and presumably is the classical representation of the
operator driving the perturbation away from the CFT point in \cite{Ahn1990,Zamolodchikov:1991vg, Ravani}.
The parafermions have fractional conformal dimension given by \eqref{dk1k2},
so that the perturbation is relevant.  Hence we expect that the $\beta$-function for $\l$ that we shall later compute,
will be linear for small $\l$.\footnote{Parafermion
bilinears dressed with other fields driving exactly marginal deformations in $\sigma$-models corresponding to exact
CFTs have been used in \cite{Petropoulos:2006py}.}
The situation is similar to the one encountered for the $\l$-deformation of the $SU(2)/U(1)$ coset CFT constructed in
\cite{Sfetsos:2013wia}, where in that case a
bilinear in the Abelian parafermions of \cite{Bardakci:1990ad} was driving the deformation.

\subsection{A non-trivial symmetry}

Similar to the case of the $\l$-deformed models \cite{Sfetsos:2013wia} and its generalizations in \cite{Georgiou:2016zyo}
and in \cite{Georgiou:2016urf}, the action \eqref{resact} has a non-trivial symmetry.
Namely, it is invariant under the transformation
\begin{equation}
\label{symmetry}
\begin{split}
&g_i\mapsto g_i^{-1}\,,\qquad k_i\mapsto-k_i\,,\qquad i=1,2\,,\\
&\l\mapsto \frac{1-(s_1-s_2)^2\l}{(s_1-s_2)^2-(1-8s_1s_2)\l}\ .
\end{split}
\end{equation}
Unlike previous works, this transformation acts non-trivially on the deformation parameter $\l$,
instead of simply inverting it.\footnote{When $k_1=k_2$ the transformation simply inverts $\l$.}
Nevertheless, this symmetry shares the $\mathbb{Z}_2$ property, i.e. when it is performed twice we get the identity.
To prove the invariance of \eqref{resact} under \eqref{symmetry} we use the transformations
\begin{equation}
D_{iab}\mapsto D_{iba}\,,\qquad J^a_{i+}\mapsto-D_{iba}\,J^b_{i+}\,,\qquad J^a_{i-}\mapsto-D_{iab}J^b_{i-}\,,\quad i=1,2\,,
\end{equation}
and also note that the Wess--Zumino terms are separately invariant. Then we work out the transformation of $\L_{12}$ defined in
\eqref{l12}, finding that
\begin{equation}
\L_{12}\mapsto\frac{4s_1s_2}{(s_1-s_2)^2 -(1- 8 s_1 s_2) \l}\,D_1^T\L_{12}\,D_2^T\
\end{equation}
and similarly for $\L_{21}$.
Using the above it is a long but straightforward computation to prove that the combined
sum of the kinetic terms of the WZW models along with the interacting pieces are invariant under \eqref{symmetry}.

The symmetry \eqref{symmetry} has two fixed points for the parameter $\l$ which are given by
\begin{equation}
\l =1  \ ,\quad \l_f= {1\ov 1-8 s_1s_2}\ .
\end{equation}
Such fixed points require special attention.
Recall that,
for the actions corresponding to the $\l$- and related deformations, the analog of \eqref{symmetry} involves
 $\l\mapsto 1/\l$
and the fixed points of the transformation are $\l=\pm 1$.
Then, a zoom in procedure for the group element around the identity has been performed and shown to
correspond to the non-Abelian T-dual of the $\sigma$-model for PCM \cite{Sfetsos:2013wia} (for $\l\to 1$)
and the corresponding
pseudodual chiral model \cite{Georgiou:2016iom} (for $\l\to -1$).
In the case at hand taking $\l=1$, leads to the
 $G/G\times G/G$ coset CFT, which is a topological model.
This issue and the associated zoom in limit will be examined in detail later in the paper.
We were not able to take a limit associated with the $\l=\l_f$ symmetry fixed point.

\subsection{Integrability}

We shall prove that the model is classically integrable by recasting its equation of motion into a Lax pair.
First we substitute \eqref{dggd} and \eqref{dggd2} into \eqref{eqg1g2} and \eqref{eqg1g22} finding that
\begin{equation}
\label{eqint1}
\begin{split}
& 4\l s_1 \del_+A_{1-} - (1-\l + 4 s_1 \l)\del_- A_{1+} + (1-\l)\del_- A_{2+}
\\
&\hskip 3 cm = (1-\l+ 4s_1 \l) [ A_{1+},A_{1-}]  -  (1-\l)[A_{2+},A_{1-}]\ ,
\\
&
(1-\l+4s_1\l) \del_+A_{1-} -4 s_1 \l \del_- A_{1+} -(1-\l) \del_+A_{2-}
\\
& \hskip 3 cm  ={ (1-\l + 4s_1 \l)} [ A_{1+},A_{1-}]  - (1-\l)[A_{1+},A_{2-}]\ 
\end{split}
\end{equation}
and then
\begin{equation}
\label{eqint2}
\begin{split}
& 4\l s_2 \del_+A_{2-} - (1-\l + 4 s_2 \l)\del_- A_{2+} + (1-\l)\del_- A_{1+}
\\
&\hskip 3 cm = (1-\l+ 4s_2 \l) [ A_{2+},A_{2-}]  -  (1-\l)[A_{1+},A_{2-}]\ ,
\\
&
(1-\l+4s_2\l) \del_+A_{2-} -4 s_2 \l \del_- A_{2+} -(1-\l) \del_+A_{1-}
\\
& \hskip 3 cm  ={ (1-\l + 4s_2 \l)} [ A_{2+},A_{2-}]  - (1-\l)[A_{2+},A_{1-}]\  .
\end{split}
\end{equation}
Not all the above equations are independent, since the difference of the two equations in \eqref{eqint1} and the difference
of those in \eqref{eqint2} are the same and given by
\begin{equation*}
\del_+ A_{1-} + \del_- A_{1+} - \del_+ A_{2-} - \del_- A_{2+}  + [A_{1+},A_{2-}]-  [A_{2+},A_{1-}]= 0 \ .
\end{equation*}
This fact is related to the existence of the constraint \eqref{sumsum}. It is convenient to define the
combinations for the gauge fields
\begin{equation}
\mathcal{A}_{\pm}=\frac12\,\left(A_{1\pm}+A_{2\pm}\right)\ ,
\quad \mathcal{B}_{\pm}=\frac12\,\left(A_{1\pm}-A_{2\pm}\right)\ .
\end{equation}
Then after some algebraic manipulations the three independent equations in \eqref{eqint1} and \eqref{eqint2} can be
can be recast as
\begin{equation}
\begin{split}
\label{eomApmcoset}
&\partial_\pm \mathcal{B}_{\mp}=[\mathcal{A}_{\pm},\mathcal{B}_{\mp}]\pm \a [\mathcal{B}_{+},\mathcal{B}_{-}] \ ,\\
&\partial_+ \mathcal{A}_{-}-\partial_- \mathcal{A}_{+}=
[\mathcal{A}_{+},\mathcal{A}_{-}]+\b [\mathcal{B}_{+},\mathcal{B}_{-}]\ ,
\end{split}
\end{equation}
with coefficients given by
\begin{equation}
\label{albe}
\a=-\frac{(s_1-s_2)(1-\l)}{1-(1-8s_1s_2)\l}\,,\qquad \b=\frac{1+\l-2(1-4s_1s_2)\l^2}{\l(1-(1-8s_1s_2)\l)}\ .
\end{equation}
Then the Lax form follows as
\begin{equation}
\label{integrabb}
\begin{split}
&{\mathcal L}_\pm=\mathcal{A}_{\pm}+\zeta_\pm \mathcal{B}_{\pm}\,,\quad \zeta_\pm=z^{\pm1}\sqrt{\a^2+\b}+\a\ ,
\\
&\partial_+{\mathcal L}_--\partial_+{\mathcal L}_+=[{\mathcal L}_+,{\mathcal L}_-]\ ,
\end{split}
\end{equation}
where $z\in\mathbb{C}$ is the spectral parameter. Note that for the case of equal level $k_1=k_2$, the
parameter $\a=0$ and $\b=1/\l$. Then the result for the Lax pair found for the $\l$-deformations of coset CFTs
corresponding to symmetric spaces \cite{Hollowood:2014rla} follows.

\section{Various limits of the effective action}
\label{limits.section}

The action \eqref{resact} admits three different limits involving the
parameters $(\l,k_1,k_2)$ as well as the group elements $g_{1,2}$.

\subsection{Non-Abelian T-dual of the WZW model for $G_{k_1}$}

We shall take one of the levels, say $k_2$, to infinity and similarly we zoom in for the group element $g_2$,
around the identity.  Specifically, consider the limit
\begin{equation}
\label{dhuh}
g_2= \mathbb{I} + i k_1 {v\ov k_2}\ , \qq k_2\to \infty \ ,
\end{equation}
where $k_1$ was inserted simply for convenience since that simplifies
the final result and $v=v_a t^a$.
When this limit is taken into the action \eqref{resact} one obtains
\begin{equation}
\label{acctas2}
S_{k_1}(g_1,v)= S_{k_1}(g_1) + {k_1\ov \pi} \int \text{d}^2\s \
(J_{1+} +  \del_+ v)(\mathbb{I}+f - D_1^T)^{-1}
 (J_{1-} +  \del_- v)\  ,
\end{equation}
where the matrix elements are
\begin{equation}
\label{fdef}
f_{ab}=f_{abc} v^c\ ,
\end{equation}
and we also recall that  we should gauge fix $\dim G$ parameters among those in $g_1$ and the $v$'s.
This action is independent of $\l$ and in fact it is nothing but the non-Abelian of the WZW model for $G_{k_1}$
\cite{Giveon:1993ai}.\footnote{One way to see
that, is to first realize from \eqref{eqpl} and \eqref{eqmi} that in the limit
\eqref{dhuh} we have that $A_{1\pm}=A_{2\pm}$. Then, one easily sees that in \eqref{acct1} the last term vanishes while the
rest of the terms form, after carefully taken  the limit and some algebraic manipulations, the starting point for performing 
a non-Abelian transformation on the WZW model  for $G_{k_1}$ with the result given by \eqref{acctas2}.}
This kind of non-Abelian T-duality is distinct from that on a PCM for $G$.
The action \eqref{acctas2} is canonically equivalent to the WZW action for $G$ \cite{Sfetsos:1996pm}.
As such the metric and antisymmetric tensor one reads from it, when supplemented with the dilaton field
$$
\text{e}^{-2\Phi} = \det(\mathbb{I}+f - D_1^T)\ ,
$$
solve the corresponding one-loop $\beta$-function equations. This will be verified below by showing that in the limit
$k_2\to \infty$ the $\beta$-function for the deformation parameter $\l$ vanishes.
In \cite{Ahn1990,Zamolodchikov:1991vg,Ravani} this limit was argued to correspond to the WZW model based on TBA considerations.
This is consistence with our findings since the WZW model action for $G$ is canonically equivalent to \eqn{acctas2} as mentioned above.

\no
The result \eqref{acctas2} of this limiting procedure is not totally surprising.
It has been known that the action \eqn{cftww} corresponding to the $G_{k_1}\times G_{k_2}/G_{k_1+k_2}$ coset CFTs,
in the limit $k_2\to \infty$ corresponds to the non-Abelian T-dual action \eqref{acctas2}, found in \cite{Sfetsos:1994vz}.
In the regime \eqref{dhuh} the same result is obtained
for any value of $\l\neq 1$ in the deformed action \eqref{resact} as well. One may wonder what is different in the case
of the $\l$-deformations \cite{Sfetsos:2013wia}, where a limit analogous to \eqref{dhuh}, but in addition with $\l$ approaching unity,
led to the non-Abelian  T-dual of the PCM for $G$ with respect to the left symmetry action it.
The essential difference is that one has to take both
levels to infinity, i.e. $k_1\to  \infty$, as well in order to obtain the non-Abelian T-dual of the PCM for
$G\times G/G$. This is explicitly shown below.

\subsection{The non-Abelian T-dual of $G\times G/G$ PCM}
\label{ZamPCM}

Consider now a limit involving both levels sent to infinity and both groups elements expanded around unity.
Specifically we let
\begin{equation}
\label{jsgh2}
g_{i}=\mathbb{I} +  2 i {v_i\ov k_i}\ ,\quad \l=1-2{\k^2\ov k}\ ,\quad
k_i\to \infty\ ,\quad i=1,2\ .
\end{equation}
Then, we obtain from \eqref{resact} that
\begin{equation}
\label{khdn}
S_{\k^2}(v)= {2\ov \pi} \int \text{d}^2\s\, \del_+ v_1 \S_{21}^{-1}\big(\k^2 (\del_- v_1 + \del_- v_2)
+ 4f_2 \del_-v_1\big) + (1 \leftrightarrow 2)\ ,
\end{equation}
where
$$
\S_{21}=\k^2(f_1 + f_2) + 4 f_2 f_1\ ,
$$
with $\S_{12}$ given by interchanging the indices $1$ and $2$ and the $f_i$'s defined as in \eqref{fdef}
by replacing the $v$'s with the $v_i$'s accordingly.
This is the non-Abelian T-dual of the PCM for the coset space $G\times G/G$. In order to see that consider
taking the limit \eqref{jsgh2} in the action \eqref{acct1}. 
After some algebraic manipulations one finds
the appropriate action, but before integrating out the gauge fields.
In \cite{Ahn1990,Zamolodchikov:1991vg,Ravani}, this limit was argued to correspond to the PCM model based on a TBA analysis.
This is in consistence with our finding since the PCM action and its non-Abelian T-dual \eqn{khdn} are 
canonically equivalent \cite{Curtright:1994be,Lozano:1995jx}.

\subsection{A non-Abelian type T-dual of the $\l$-deformed $\sigma$-models}
\label{nonabelLambda}

Finally we consider the case where the limit \eqref{dhuh} is also taken but simultaneously $\l$ approaches unity.
The level $k_1$ still remains finite.
Specifically,  let us consider the limit
\begin{equation}
\label{dhuh2}
g_2= \mathbb{I} + i  {k_1\ov k_2} v\ ,\qq \l = 1-{k_1 \ov k_2}\k^2\ , \qq k_2\to \infty \ .
\end{equation}
When this is taken in \eqref{resact}, we find the result
\begin{equation}
\label{hgkje}
\begin{split}
&
S_{k_1,\k^2}(g_1,v) =  S_{k_1}(g_1) + {k_1\ov \pi} \int \text{d}^2\s \Big[ J_{1+}\S^{-1}
 (J_{1-} +  \del_- v + 4\k^{-2} f J_{1-})\,
\\
& \qq\qq\qq\qq
+ \del_+ v \widetilde\S^{-1}  (J_{1-} +  \del_- v + 4\k^{-2} (\mathbb{I}-D_1^T) \del_-v\Big]\ ,
\end{split}
\end{equation}
where
$$
\S = \mathbb{I}+f - D_1^T + 4\k^{-2} f(\mathbb{I}-D_1^T)\ ,\quad
\widetilde\S = \mathbb{I}+f - D_1^T + 4\k^{-2} (\mathbb{I}-D_1^T)f\ .
$$
Note that \eqref{hgkje} reduces to \eqref{acctas2} for $\k\to \infty$. The reason is that, this limit effectively
moves the parameter $\l$ away from unity so that the limit \eqref{dhuh2} reduces to that in \eqref{dhuh}.

The limit \eqref{dhuh2} suggests that the above result corresponds to some kind of non-Abelian
T-dual limit.
However, the construction has some notable differences as compared with the traditional non-Abelian T-duality transformation.
Indeed, considering the
limit \eqref{dhuh2} in the action \eqref{acct1} before the gauged fields are integrated out we obtain that
\begin{equation}
\label{acct13}
\begin{split}
&
S_{k_1,\k^2}(g_1,A_\pm) = S_{k_1}(g_1)
 +{k_1\ov \pi} \int \text{d}^2\s\ \Tr \big(A_{1-} \del_+ g_1 g_1^{-1} - A_{1+} g_1^{-1} \del_- g_1
 \\
 &\phantom{xxxxxxxxx} + A_{1-} g_1 A_{1+} g_1^{-1}  -A_{1+} A_{1-} - iv F_{2+-} \big)
 - {k_1 \k^2\ov \pi } \int \text{d}^2\s\ {\rm Tr} \big(\mathcal{B}_+\mathcal{B}_-\big) \ .
\end{split}
\end{equation}
Integrating out the $A_{i\pm}$'s we obtain of course \eqn{hgkje}. However,
integrating out the Lagrange multiplier term $v$ forces
the gauge field $A_{2\pm}$ to be a pure gauge. Choosing $A_{2\pm}=0$, we remain with the action
\begin{equation}
\label{acct14}
\begin{split}
&
S_{k_1,\l}(g_1,A_\pm) = S_{k_1}(g_1)
 +{k_1\ov \pi} \int \text{d}^2\s\ \Tr \big(A_{1-} \del_+ g_1 g_1^{-1} - A_{1+} g_1^{-1} \del_- g_1
 \\
 &\phantom{xxxXXxxxxxx} + A_{1-} g_1A_{1+} g_1^{-1}  -\l_0^{-1}A_{1+} A_{1-} \big)\ ,
\end{split}
\end{equation}
where $\l_0^{-1}=1+\frac{\k^2}{4}$. 
This is nothing but the action, before integrating out the remaining non-propagating fields
$A_{1\pm}$'s, for the usual $\l$-deformed $\sigma$-models \cite{Sfetsos:2013wia}. 
Since non-Abelian T-duality is generally speaking a canonical transformation this equivalence will show up in the RG flow equation for $\k^2$
(equivalently $\l_0$) that we shall compute in the next section.

Finally, we note that by performing the traditional non-Abelian T-duality transformation to the action for $\l$-deformation for the global 
invariance having a vector action on the group element, the result is \eqn{hgkje}.

\section{Renormalization group flows}
\label{RGflows.section}

The scope of this section is to compute the $\beta$-function of the coupling constant $\l$.
We shall use a method developed in the present context in \cite{Appadu:2015nfa} and in \cite{Georgiou:2017aei}.
To proceed we need to determine a specific background solution
and evaluate its quantum fluctuations.
The equations of motion are given by \eqref{eomApmcoset}. In addition we fix the residual gauge through
the covariant gauge fixing condition
\begin{equation}
\label{cosetgauge}
\partial_+\mathcal{A}_{-} +\partial_-\mathcal{A}_{+}=0\,.
\end{equation}
At first we specify a particular background solution by parameterizing the group elements as
\begin{equation}
g_i=\text{e}^{\sigma^\mu\Theta_{i\mu}}\,,\quad i=1,2\ ,\quad \m=+,-\ ,
\end{equation}
where the $\Theta_{i\mu}$'s, are constant commuting elements in the Lie algebra of the group $G$.
Next, we set $\mathcal{A}_{\pm}=0$ so that we project to the coset $G\times G/G$.  Then, we evaluate the gauge fields $\mathcal{B}_{\pm}$ on this background 
\begin{equation}
\mathcal{B}_{\pm}=\pm\frac{\l}{1-\l}\left(s_1\Theta_{1\pm}-s_2\Theta_{2\pm}\right)\ .
\end{equation}
which satisfy the equations of motion \eqref{eomApmcoset} and the gauge fixing \eqref{cosetgauge}.

The Lagrangian density for this background is easily found to be 
\begin{equation}
\begin{aligned}
{\mathcal L}^{(0)}&=-\frac{k_1}{2\pi}\Theta_{1+}\Theta_{1-}-\frac{k_2}{2\pi}\Theta_{2+}\Theta_{2-}\\
&-\frac{k_1+k_2}{\pi}\frac{\l}{1-\l}
\left(s_1\Theta_{1+}-s_2\Theta_{2+}\right)\left(s_1\Theta_{1-}-s_2\Theta_{2-}\right)\ .
\end{aligned}
\end{equation}
Next we vary the equations of motion \eqref{eomApmcoset} and the gauge fixing condition \eqref{cosetgauge} obtaining that
\begin{equation}
\left(\begin{array}{cccc}
\partial_- +\a \tilde{\mathcal{B}}_{-}& -\a \tilde{\mathcal{B}}_{+} & 0 & -\tilde{\mathcal{B}}_{+}	\\
-\a \tilde{\mathcal{B}}_{-} & \partial_+ +\a \tilde{\mathcal{B}}_{+} & -\tilde{\mathcal{B}}_{-} & 0	\\
-\b \tilde{\mathcal{B}}_{-} & \b \tilde{\mathcal{B}}_{+} & -\partial_- & \partial_+	\\
0 & 0 & \partial_- & \partial_+
\end{array}\right)
\left(\begin{array}{c}
\d \mathcal{B}_{+}\\
\d \mathcal{B}_{-}\\
\d \mathcal{A}_{+}\\
\d \mathcal{A}_{-}
\end{array}\right)=0\,,
\end{equation}
with
$\left(\tilde{\mathcal{B}}_{\pm}\right)_{ab}=i f_{abc}\, \mathcal{B}^c_{\pm}$ and
$\a,\b$ where defined in \eqref{albe}.
To evaluate the one-loop effective Lagrangian, we Wick rotate to Euclidean space and then
we integrate out the fluctuations
in the Gaussian path integral. The result in momentum space reads
\begin{equation}
-{\mathcal L}^\text{eff}_\text{E}={\mathcal L}^{(0)}+
\int^\mu\frac{\text{d}^2p}{(2\pi)^2}\ln\det{\mathcal D}^{-1/2}\,,\quad \text{d}^2p=\text{d}p_1\text{d}p_2\,,
\end{equation}
where $\mu$ is a cutoff scale and
\begin{equation}
{\mathcal D}=\left(\begin{array}{cccc}
p_- +\a \tilde{\mathcal{B}}_{-}& -\a \tilde{\mathcal{B}}_{+} & 0 & -\tilde{\mathcal{B}}_{+}	\\
-\a \tilde{\mathcal{B}}_{-} & p_+ +\a \tilde{\mathcal{B}}_{+} & -\tilde{\mathcal{B}}_{-} & 0	\\
-\b \tilde{\mathcal{B}}_{-} & \b \tilde{\mathcal{B}}_{+} & -p_- & p_+	\\
0 & 0 & p_- & p_+
\end{array}\right)\,.
\end{equation}
Working along the lines of  \cite{Appadu:2015nfa, Georgiou:2017aei}, after some algebra we
obtain
\begin{equation}
-{\mathcal L}_E^\text{eff}={\mathcal L}^{(0)}-\frac{c_G}{\pi}\left(\alpha^2+\beta\right)\frac{\l^2}{(1-\l)^2}
\left(s_1\Theta_{1+}-s_2\Theta_{2+}\right)\left(s_1\Theta_{1-}-s_2\Theta_{2-}\right)\ln\mu\,.
\end{equation}
The one-loop $\beta$-function is derived by demanding that the effective action in independent of the cutoff scale $\mu$.
To leading order in the large level expansion we obtain that
\begin{equation}
\label{betafunction}
\beta_\l=\frac{\text{d}\l}{\text{d}\ln\mu^2}=-\frac{c_G\l(1-\l_1^{-1}\l)(1-\l_2^{-1}\l)(1-\l^{-1}_3\l)}
{2(k_1+k_2)(1-\l_f^{-1}\l)^2}\,,
\end{equation}
where
\begin{equation}
\label{betafunctiondef}
\l_1=\frac{1}{s_2-3s_1}\,,\quad \l_2=\frac{1}{s_1-3s_2}\,,\quad
 \l_3=\frac{1}{(s_1-s_2)^2} \ .
\end{equation}
The $\beta$-function is symmetric in exchanging $k_1$ with $k_2$ and it
is invariant under the symmetry \eqref{symmetry}, under which the points \eqref{betafunctiondef}
map to each other as
\begin{equation}
\l_1\mapsto\l_2\,,\quad\l_2\mapsto\l_1\,,\quad\l_3\mapsto0\,.
\end{equation}

\subsection*{Properties of the $\beta$-function}

\begin{enumerate}

 \item
 The $\beta$-function \eqn{betafunction} has four fixed points at $\l=(0,\l_{1,2,3})$, where $\l_{1,2,3}$
were defined in \eqref{betafunctiondef}.
Near $\l=0$ we obtain the  $G_{k_1}\times G_{k_2}/G_{k_1+k_2}$ coset CFTs
perturbed by a parafermion bilinear, as in \eqref{paraf}.
This is in agreement with the behavior of the $\beta$-function for small $\l$ given by
\begin{equation}
\b_\l\simeq-\frac{c_G\l}{2(k_1+k_2)}+{\cal O}(\l^2)\ .
\end{equation}
Hence, the operator driving the perturbation is relevant and has scaling dimension
\begin{equation}
\D=2-\frac{c_G}{k_1+k_2}\, .
\end{equation}
This is in agreement with \eqref{dk1k2} for large $k_1$ and $k_2$.

\no
To analyze the fixed points $\l_1$ and $\l_2$ we choose without loss of generality that $k_1>k_2$.
We find that at $\l=\l_1$ the action \eqn{resact} becomes that in
\eqn{cftww} with the replacement $k_1\mapsto k_1-k_2$, $k\mapsto k_1$ and for the group element $g_2\mapsto g_1g_2$.
Hence,
\begin{equation}
\label{CFTinterp}
\l=\l_1:\qq   \frac{G_{k_1-k_2}\times G_{k_2}}{G_{k_1}}\, ,
\end{equation}
which is a unitary coset CFT. As we have taken $k_1>k_2$, the fixed point $\l_1$ is negative and there is an RG flow from
the UV fixed point at $\l=0$, towards the IR fixed point at $\l=\l_1$.
Obviously, the central charges at the ends of the flow are in agreement with the $c$-theorem of \cite{Zamolodchikov:1986gt}.
The $\beta$-function for $\l$ near $\l_1$ reads
\begin{equation}
\b_\l\simeq\frac{c_G(\l-\l_1)}{2(k_1-k_2)}+{\cal O}(\l-\l_1)^2\ .
\end{equation}
Hence, the operator driving the perturbation has anomalous dimension $\frac{c_G}{k_1-k_2}$. The $\beta$-function starts 
positive which is in line with 
$\l=\l_1$ being the IR fixed point.

\no
For $\l=\l_2$ one similarly finds the theory \eqn{CFTinterp} but with $k_1$ and $k_2$ interchanged. However, this
corresponds to a non-unitary coset CFT and is not of interest.
Regarding the point $\l=\l_3$,
it is always bigger than one, which is a singular point of the action, and so it is continuously disconnected from 
the RG flow initiating at $\l=0$.

\item
When $k_1=k_2$, the $\beta$-function \eqref{betafunction} drastically simplifies to the standard
expression for symmetric spaces found with different methods in \cite{Itsios:2014lca,Sfetsos:2014jfa} and \cite{Appadu:2015nfa}
\begin{equation}
\label{GHsymmetric}
\b_\l=-\frac{c_G\l}{4k_1}\,.
\end{equation}
The complexity of \eqref{betafunction} when $k_1\neq k_2$ is explained by the Dirac-bracket algebra of the operator driving the
perturbation. The parafermionic algebra \eqref{DBcosetalgebra} (see appendix \ref{Para.append} for details of the derivation),
was found in \cite{Bardakci:1990ad}
\begin{equation}
\label{paraalgebra}
\begin{aligned}
\{\Psi_\pm^{a}(\s),\Psi_\pm^{b}(\s')\}_\text{D.B.}=&
\pm\frac{\d_{ab}\d'_{\s\s'}}{2(k_1+k_2)}
-\frac{k_1-k_2}{4(k_1+k_2)^2}f_{abc}\Psi^{c}_\pm(\s)\d_{\s\s'}\\
&\pm\frac{1}{4(k_1+k_2)}f_{aec}f_{brc}\Psi_\pm^{e}(\s)\Psi_\pm^{r}(\s')\varepsilon_{\s\s'}\,,
\end{aligned}
\end{equation}
where $\d_{\s\s'}= \d(\s-\s')$ is the usual $\d$-function and
 $\varepsilon_{\s\s'}=\varepsilon(\s-\s')$ is the antisymmetric step function, so that $\varepsilon'_{\s\s'}=2\delta_{\s\s'}$. In
addition
\begin{equation*}
\Psi_+=s_1(D_+g_1 g_1^{-1}+A_{1+}-A_{1-})\,,\quad
\Psi_-=-s_1(g_1^{-1}D_-g_1+A_{1+}-A_{1-})\,.
\end{equation*}
A comment is in order regarding the expansion around $\l=0$. At that point
the above expression of $\Psi$'s coincides with \eqref{defparaf}
and so the perturbation \eqref{paraf} is a bilinear of parafermions that satisfy the algebra \eqref{paraalgebra}.

The appearance of the second term in \eqref{paraalgebra} for $k_1\neq k_2$, resembles the analogue formula for general
coset spaces $G_k/H_k$ \cite{Bardacki:1990wj}
\be
\begin{split}
\label{paraalgebra.GH}
&\{\Psi_\pm^{\a}(\s),\Psi_\pm^{\b}(\s')\}_\text{D.B.}=
\pm\frac{2}{k}\d_{\a\b}\d'_{\s\s'}+\frac{2}{k}f_{\a\b\g}\Psi_\pm^{\g}(\s)\d_{\s\s'}
\\
&\qq\qq\qq \pm \frac{1}{k}f_{\a\g c}f_{\b\d c}\Psi_\pm^{\g}(\s)\Psi_\pm^{\d}(\s')\varepsilon_{\s\s'}\ ,
\\
&\Psi_+=D_+g g^{-1}+A_{+}-A_{-}\,,\quad
\Psi_-=-g^{-1}D_-g-A_{+}+A_{-}\,,
\end{split}
\ee
where the Greek and Latin indices correspond to generators in the coset and subgroup respectively.
For symmetric spaces, the second term drops out since then $f_{\a\b\g}=0$, alike \eqref{paraalgebra}
for $k_1=k_2$, and the $\beta$-function is given by \eqref{GHsymmetric}.

\item
The $\beta$-function possesses two interesting expansions around the fixed
points of the symmetry $\l=1$ and $\l=\l_f=(1-8 s_1s_2)^{-1}$, when $k_{1,2}\gg1$ uncorrelated with
$|k_1-k_2|=n$, where $n$ is a finite number.
In particular:
\begin{itemize}
 \item
Expanding near $\l=1$, we obtain the $\beta$-function of the PCM for the overall coupling $\k^2$
\begin{equation}
\frac{\text{d}\k^{2}}{\text{d}\ln\mu^2}=\frac{c_G}{4}\,,\quad \l=1-\frac{2\k^{2}}{k_1+k_2}\,,\quad k_{1,2}\gg1\ .
\end{equation}
The action corresponding to this limit was derived in \eqn{khdn} and is the non-Abelian T-dual of the PCM for $G\times G/G$.
The $\beta$-function for $\k^2$ is the same as that for the corresponding PCM, since 
the two models are related by a canonical transformation.

\item
Around $\l=\l_f$, we obtain the $\beta$-function of the non-critical WZW with $n$ being the coupling of the WZ term
\cite{Witten:1983ar}
\begin{equation}
\frac{\text{d}\k^{2}}{\text{d}\ln\mu^2}=\frac{c_G}{4}\left(1-n^2\k^{-4}\right),\quad
\l\simeq\l_f+\frac{2\k^{2}}{k_1+k_2}\,,\quad k_{1,2}\gg1\,,
\end{equation}
Unlike the previous case we were unable to show that this limit is also realized at the action level by taking 
an appropriate limit in \eqn{resact}.

\end{itemize}
The above results are in align with the predictions in \cite{Ahn1990, Zamolodchikov:1991vg, Ravani} using TBA techniques.

\item
An interesting variation of the above limiting expansion around $\l=1$ is to consider one of the levels going to infinity whereas
keeping the other one finite, i.e. $k_2\gg1$, while keeping $k_1$ finite. We find that
\begin{equation}
\frac{\text{d}\k^{2}}{\text{d}\ln\mu^2} = \frac{2c_G}{k_1}\left(\frac{\k^2+4}{\k^2+8}\right)^2,
\quad \l\simeq1-\frac{k_1}{k_2}\k^2\,,\quad k_2\gg1\,.
\end{equation}
After the replacement $ \l_0^{-1}=1+\frac{\k^2}{4}$ (we use $\l_0$ instead of $\l$ to avoid confusion)
this can equivalently expressed as the $\beta$-function for the $\lambda$-deformed model at level $k_1$ found in
\cite{Itsios:2014lca}
\begin{equation}
\frac{\text{d}\l_0}{\text{d}\ln\mu^2}=-\frac{c_G\l_0^2}{2k_1(1+\l_0)^2}\ .
\end{equation}
This limit when taken at the level of the action gave \eqn{hgkje}. As previously noted in 
section \ref{nonabelLambda}, the action \eqn{hgkje} is the non-abelian T-dual of the $\lambda$-deformed model at level $k_1$. 
Therefore the models have the same $\b$-function as they are related by a canonical transformation.

\item
It would be important to derive the $\beta$-function \eqref{betafunction} using gravitational methods. We were able to do so for the 
case with $G=SU(2)$. In particular, we used the background corresponding to the $SU(2)_{k_1}\times SU(2)_{k_2}/SU(2)_{k_1+k_2}$ 
$\l$-deformed model found in \cite{Sfetsos:2014cea}. This background has
zero antisymmetric tensor and metric and dilaton given by
\begin{equation}
\begin{aligned}
\label{dsdefk1k2}
& \text{d}s^2  = \frac{2(k_1+k_2)}{(1-\l)\Lambda}
\Big( \Omega_{\a\a} \text{d}\a_0^2  +  \Omega_{\b\b} \text{d}\b_0^2+   \Omega_{\g \g} \text{d}\g^2\\
&\phantom{xxxxxx}
 +2 \Omega_{\a \b}\text{d}\a_0 \text{d}\b_0 + 2 \Omega_{\a\gamma} \text{d}\a_0 \text{d}\g
+ 2 \Omega_{\b\gamma} \text{d}\b_0 \text{d}\g  \Big)\,,\\
&\text{e}^{-2\Phi}=\L\,,\quad \L=(1-\a_0^2)(1-\b_0^2)-\g^2\,,
\end{aligned}
\end{equation}
with
\begin{equation}
\begin{aligned}
 & \Omega_{\a\a} =   (1+ r)^{-2} Z^{-1} \left( Z^2 - \left(Z^2-  (1-\l)^2(1+r^{-1})^2 \right)\b_0^2 \right)\,,
 \\ & \Omega_{\b\b}  =  (1+ r^{-1})^{-2} Z^{-1} \left( Z^2 - \left(Z^2-  (1-\l)^2(1+r)^2\right)\a_0^2 \right)\,, \\
&  \Omega_{\g\g} =    (1-\l)^{2} Z^{-1}\,,\\
&\Omega_{\a \b} =  (1-\l)^2 Z^{-1} \a_0 \b_0 + r (1+r)^{-2} Z \g\,,\\
& \Omega_{\a\g}   = -r^{-1} (1-\l)^2 Z^{-1} \b_0\,, \quad
 \Omega_{\b\g}   = -r (1-\l)^2 Z^{-1} \a_0 \,,
\end{aligned}
\end{equation}
and
\begin{equation*}
r = \frac{k_2}{k_1} \,,\quad Z =  8 \l + (1-\l) r^{-1}(1+r)^2 \,.
\end{equation*}
To compute the corresponding $\beta$-function we employ the background field expansion \cite{honer,Friedan:1980jf,Curtright:1984dz}
\begin{equation}
\frac{\text{d}g_{\mu\nu}}{\text{d}\ln\mu^2}=R_{\mu\nu}+\nabla_\mu\xi_\nu+\nabla_\nu\xi_\mu\ ,
\end{equation}
where $\xi^\m$ is a vector corresponding to possible diffeomorphisms along the RG flow.
The result of the computation is precisely the $\beta$-function \eqref{betafunction} with $c_G=4$ (appropriate for the
quadratic Casimir in the adjoint representation of $SU(2)$) and $\xi_\mu=\partial_\mu\Phi$.

\item
It was shown in \cite{Sfetsos:2014cea} that the above target space \eqref{dsdefk1k2} can be embedded in type-IIB supergravity.
In particular, the metric and the dilaton are
supported by a three-form $F_3$, given in (A.9) and (A.10) of \cite{Sfetsos:2014cea}, with an overall
real coefficient labelled by $\mu$ (not be confused with the cutoff scale in the RG flow equations above)
\begin{equation}
\mu^2=-\frac{32s_1s_2}{(1-\l)(1-\l_f^{-1}\l)}\beta_\l\ .
\end{equation}
We remark that this is invariant under the symmetry \eqref{symmetry}.
Initiating an RG flow from $\l=0$, regularity of the solution and the positivity of $\mu^2$ retain $\l\in[0,1)$
and disregard for $k_1>k_2$ the domain $\l\in[\l_1,0]$ since then $\mu^2<0$.

\end{enumerate}

\section{Outlook}

In this work we investigated $\l$-deformations of the $G_{k_1} \times G_{k_2}/G_{k_1+k_2}$ coset CFTs,
based on a semi-simple group $G$ and characterized also by two different levels $k_1,k_2$.
These models, whose action is \eqref{resact}, have some very attractive features.
They are invariant under the non-trivial symmetry \eqref{symmetry} and prove to be classically integrable, as their equations of motion can be written in the Lax form \eqref{integrabb}. It will be interesting to prove integrability in the strong sense, as it was done for the $\l$-deformations of WZW models in \cite{Itsios:2014vfa}. This requires a generalization of the Maillet brackets \cite{Maillet2} but now in the
presence of terms containing the antisymmetric step function $\varepsilon_{\s\s'}$.
We have computed the exact $\b$-function in \eqref{betafunction} and shown that it
possesses a non-trivial fixed point, unlike the symmetric case with $k_1=k_2$. Hence, there is a smooth RG flow
from the $G_{k_1} \times G_{k_2}/G_{k_1+k_2}$ coset CFTs  in the UV to  $ G_{k_1-k_2}\times G_{k_2}/G_{k_1}$ coset CFTs in the IR. The flow is driven by parafermion bilinears \eqref{paraf}, whose conformal dimension is given in \eqref{dk1k2}.
These satisfy the parafermionic algebra \eqref{paraalgebra} whose structure explains the difference between the case of equal and
unequal levels. In that respect the models constructed here have certain similarities with the left-right asymmetric $\l$-
deformations of WZW models perturbed by current bilinears having two different levels and which also possess a non-trivial
IR fixed point \cite{Georgiou:2016zyo,Georgiou:2017jfi}.

\no
Our models possess various interesting limits when one or both levels $k_1$ and $k_1,k_2$ are taken to infinity. These are in resonance
with finding of previous works which used thermodynamic Bethe ansatz techniques. They can be embedded in type-IIB SUGRA, when $G=SU(2)$ as it was shown in \cite{Sfetsos:2014cea}.
There are several open directions which need to be further pursued.
In particular, it would be very interesting to derive the $\beta$-function \eqref{betafunction} when
$k_1\neq k_2$ using CFT techniques from the OPEs for the parafermions $\Psi_\pm$,
obeying the algebra \eqref{paraalgebra}. In fact this may be pursued $\l$-deformed general coset spaces $G_k/H_k$
and the analogous parafermionic algebra \eqref{paraalgebra.GH}. Given the experience with $\l$-deformations of WZW models 
we expect that the symmetry \eqn{symmetry} and some perturbative information should be enough to 
reevaluate the exact $\beta$-function \eqn{betafunction} and moreover compute the anomalous dimension of operators.

\appendix

\section{The parafermionic algebra}
\label{Para.append}

The purpose of this appendix is to revisit the parafermionic algebra for the
$G_{k_1}\times G_{k_2}/G_{k_1+k_2}$ coset CFTs, originally found in \cite{Bardakci:1990ad}.
In our case the coset parafermions are given by
\begin{equation}
\label{DBcosetgenerators}
\Psi_\pm=\frac12\,\left(s_1{\mathcal J}_{1\pm}-s_2{\mathcal J}_{2\pm}\right)\,,
\end{equation}
where ${\mathcal J}_{i\pm}^a$ satisfy a set of commuting current algebras \cite{Bowcock:1988xr}
\begin{equation}
\label{Bowcock}
\begin{split}
&\{{\mathcal J}_{i\pm}^a,{\mathcal J}_{i\pm}^b\}=\frac{2}{k_i}\,
\left(f_{abc}{\mathcal J}_{i\pm}^c\d_{\s\s'}\pm\delta_{ab}\d'_{\s\s'}\right)\,,\quad i=1,2\,,\\
&{\mathcal J}_{i+}=D_+g_i g_i^{-1}+A_{i+}-A_{i-}\,,\quad
{\mathcal J}_{i-}=-g_i^{-1}D_-g_i-A_{i+}+A_{i-}\,.
\end{split}
\end{equation}
Next we restrict ourselves to the coset $G_{k_1} \times G_{k_2}/G_{k_1+k_2}$,
by enforcing the constraints on the subgroup
\begin{equation}
\label{constaintJ}
{\mathcal H}_{\pm}=\frac12\,\left(s_1{\mathcal J}_{1\pm}+s_2{\mathcal J}_{2\pm}\right)\approx0\,,
\end{equation}
or equivalently through \eqref{sumsum} and \eqref{Bowcock}, in terms of gauge fields
\begin{equation}
\label{constaintA}
s_1(A_{1+}-A_{1-})+s_2(A_{2+}-A_{2-})\approx0\,.
\end{equation}
These constraints turn out to be second class as the matrix of their Poisson brackets
\begin{equation}
{\mathcal C}^{ab}_{\pm\pm}=\{{\mathcal H}_{\pm}^{a},{\mathcal H}_{\pm}^{b}\}\approx
\pm\frac{\d_{ab}\d'_{\s\s'}}{2(k_1+k_2)}\,,
\end{equation}
is invertible with
\begin{equation}
\left({\mathcal C}^{ab}_{\pm\pm}\right)^{-1}\approx
\pm(k_1+k_2)\d_{ab}\varepsilon_{\s\s'}\,,
\end{equation}
where $\varepsilon_{\s\s'}$ was defined after \eqref{paraalgebra}.
Equipped with the above we can evaluate the non-vanishing Dirac brackets for \eqref{DBcosetgenerators} throughout
the general definition
\begin{equation}
\{\Psi_\pm^{a},\Psi_\pm^{b}\}_\text{D.B.}=\{\Psi_\pm^{a},\Psi_\pm^{b}\}-
\{\Psi_\pm^{a},{\mathcal H}_{\pm}^{c}\}\left({\mathcal C}^{cd}_{\pm\pm}\right)^{-1}
\{{\mathcal H}_{\pm}^{d},\Psi_\pm^{b}\}\,,
\end{equation}
and after some algebra we obtain
\begin{equation}
\label{DBcosetalgebra}
\begin{aligned}
\{\Psi_\pm^{a}(\s),\Psi_\pm^{b}(\s')\}_\text{D.B.}&=
\pm\frac{\d_{ab}\d'_{\s\s'}}{2(k_1+k_2)}
-\frac{k_1-k_2}{4(k_1+k_2)^2}f_{abc}\Psi^{c}_\pm(\s)\d_{\s\s'}\\
&\pm\frac{1}{4(k_1+k_2)}f_{aec}f_{brc}\Psi_\pm^{e}(\s)\Psi_\pm^{r}(\s')\varepsilon_{\s\s'}\,,
\end{aligned}
\end{equation}
where
\begin{equation*}
\Psi_+=s_1(D_+g_1 g_1^{-1}+A_{1+}-A_{1-})\,,\quad
\Psi_-=-s_1(g_1^{-1}D_-g_1+A_{1+}-A_{1-})\,.
\end{equation*}
Finally we note that our result \eqref{DBcosetalgebra} is in agreement with the findings of \cite{Bardakci:1990ad}.


\end{document}